\documentclass[12pt]{iopart}
\usepackage{graphicx}
\usepackage{tikz}
\usepackage{iopams}
\newcommand{\pdagger}{{\phantom{\dagger}}}
\newcommand{\psbar}{\bar{\psi}}

\begin{document}

\title{Quantum statistical fluctuation of energy and its novel pseudo-gauge dependence}
\author{Arpan Das}
\address{Institute of Nuclear Physics Polish Academy of Sciences, PL-31-342 Krakow, Poland}

\ead{arpan.das@ifj.edu.pl}
\vspace{10pt}

\begin{indented}
\item[]January, 2022
\end{indented}

\begin{abstract}
We discuss the quantum statistical fluctuations of energy in subsystems of hot relativistic gas for both spin-zero and spin half particles. We explicitly show the system size dependence of the quantum statistical fluctuation of energy. Our results show that with decreasing system size 
quantum statistical fluctuations increase substantially. As the consistency of the framework, we also argue that the quantum statistical fluctuations give rise to the known result for statistical fluctuation of energy in the canonical ensemble if we consider the size of the subsystem to be sufficiently large. For a spin-half particle quantum fluctuations show some interesting novel features. We show that within a small sub-system quantum statistical fluctuation of energy for spin half particles depends on the various \textit{pseudo-gauge} choices of the energy-momentum tensor. Interestingly, for sufficiently large subsystems quantum fluctuations obtained for different pseudo-gauge choices converge and we recover the canonical-ensemble formula known for statistical fluctuations of energy. Our calculation is very general and can be applied to any branch of physics whenever one deals with a thermal system. As a practical application, we argue that our results can be used to determine a coarse-graining scale to introduce the concept of classical energy density or fluid element relevant for the strongly interacting matter, in particular for small systems produced in heavy-ion collisions.
\end{abstract}

%
% Uncomment for keywords
\vspace{2pc}
\noindent{\it Keywords}: Quantum fluctuation, heavy-ion collision, pseudo-gauge dependence, small systems.
%
% Uncomment for Submitted to journal title message
%\submitto{\JPA}
%
% Uncomment if a separate title page is required
%\maketitle
% 
% For two-column output uncomment the next line and choose [10pt] rather than [12pt] in the \documentclass declaration
%\ioptwocol
%
\section{Introduction}
Relativistic heavy-ion collision experiments (RHICE), e.g., the Relativistic Heavy Ion Collider (RHIC) at the Brookhaven National Laboratory (BNL) and the Large Hadron Collider (LHC) at CERN gives us a unique opportunity to investigate the properties of strongly 
interacting matter governed by Quantum Chromodynamics (QCD), particularly in the non-perturbative regime. The physics of strongly interacting matter so produced can be broadly separated into three different stages, (i) description of the initial stage, (ii) bulk evolution of the deconfined locally thermalized plasma and, (iii) hadronization of quark-gluon plasma and freezeout of the hadrons. The relativistic dissipative hydrodynamics has been successfully used to model the bulk evolution of the strongly interacting matter. Modeling of the initial stage, bulk hydrodynamic evolution, and the freezeout prescription of hadrons constitute the standard model of the heavy-ion collisions~\cite{Romatschke:2017ejr,Gale:2013da,Jeon:2015dfa,Florkowski:2017olj}. We should emphasize that hydrodynamics is a macroscopic effective theory. Therefore from a microscopic theory point of view, the notion of ``coarse-graining" or the concept of ``fluid cell/element" is required.  Further, hydrodynamic description assumes local thermal equilibrium, where the well-defined concepts of energy density, number density, etc. over a fluid element have been used. We do not argue on the applicability or success of relativistic hydrodynamics. Rather, in this article, we try to shed light on the concepts of energy density and how well it is defined for a finite-size system. In this context, we study the quantum statistical fluctuation of the time-time (``$tt$") component of the energy-momentum tensor ($\hat{\mathcal{T}}^{tt}$) within a quantum field theoretical framework for spin-zero (scalar) as well as for spin-half (fermion) particles~\cite{Das:2021acta,Das:2021aar}.

Statistical fluctuations of various thermodynamic quantities become relevant in different fields of physics, as they encode possible information about phase transitions~\cite{Smoluchowski,PhysRevLett.85.2076},
dissipative phenomena~\cite{Kubo1}, formation of structures in the Early Universe~\cite{Lifshitz:1963ps,PhysRevLett.49.1110}, etc. 
Fluctuations can encode the quantum nature of the system arising due to quantum uncertainty relation or it can be inherently present in thermodynamic systems due to its statistical nature (we call such fluctuations as statistical fluctuations)~\cite{Huang:1987asp}. 
The present study combines quantum as well as statistical aspects of fluctuations, i.e. we study quantum statistical fluctuations of energy in subsystems of hot relativistic gas~\cite{Das:2021acta,Das:2021aar}. We demonstrate that for small subsystem sizes due to the quantum uncertainty relation quantum statistical fluctuations can be significantly large. On the other hand for a large system, these quantum statistical fluctuations give rise to standard statistical fluctuations for a thermodynamic system (for a more detailed discussion see Refs.~\cite{Das:2021acta,Das:2021aar}).   

Furthermore, we argue that for spin-half particles quantum statistical fluctuation of energy crucially depends on the form of the energy-momentum tensor (EMT). Note that for a quantum field theoretical system Noether theorem does not uniquely define the energy-momentum tensor (EMT). Mathematically, for any conserved energy-momentum tensor $\hat{\mathcal{T}}^{\mu\nu}$, i.e. $\partial_\mu \hat{\mathcal{T}}^{\mu\nu}=0$ one can construct an equivalent energy-momentum tensor $\hat{\mathcal{T}}^{\prime \,\mu\nu}$ by adding the divergence of an antisymmetric tensor, namely $\hat{\mathcal{T}}^{\prime \,\mu\nu} = \hat{\mathcal{T}}^{\mu\nu} + \partial_\lambda \hat{\Phi}^{\nu\mu \lambda}$~\cite{Chen:2018cts,HEHL197655,Speranza:2020ilk}. Note that if $\hat{\Phi}^{\nu\mu \lambda} = - \hat{\Phi}^{\nu\lambda \mu}$, then $\partial_\mu \hat{\mathcal{T}}^{\mu\nu}=0$ implies $\partial_\mu \hat{\mathcal{T}}^{\prime\mu\nu}=0$. Such different choices of conserved energy-momentum tensor is also known as pseudo-gauge choices and the tensor $\hat{\Phi}^{\mu\nu\lambda}$ is called the pseudo-gauge parameter. To study the pseudo-gauge dependence of quantum statistical fluctuation of energy we consider the canonical  version of the energy-momentum tensor for spin-half particle along with 
the Belinfante-Rosenfeld (BR) energy-momentum tensor~\cite{BELINFANTE1939887,BELINFANTE1940449,Rosenfeld1940}, the de Groot-van Leeuwen-van  Weert  (GLW) energy-momentum tensor~\cite{DeGroot:1980dk}, and the Hilgevoord-Wouthuysen (HW) energy-momentum tensor~\cite{HILGEVOORD19631,HILGEVOORD19651002}. Note that in the context of hyperon polarization as observed in the heavy-ion collision experiments, physical implications of pseudo-gauge transformation of the energy-momentum tensor and the spin tensor  have been extensively discussed~\cite{HEHL197655,Speranza:2020ilk,Florkowski:2018fap,Leader:2013jra,Gallegos:2021bzp,Buzzegoli:2021wlg}.

For a qualitative and quantitative estimation of quantum statistical fluctuation, we first derive a compact analytical formula for fluctuations in a subsystem of hot relativistic gas. Then we use this analytical expression of quantum statistical fluctuation for numerical estimation keeping in mind the physical situations expected in relativistic heavy-ion collisions. Throughout the manuscript we consider the Minkowski metric with mostly negative signature, i.e. $g_{\mu\nu} = \hbox{diag}(+1,-1,-1,-1)$, three vectors are represented with \textit{bold} letters and scalar product for three-vectors and Lorentz four vectors are denoted as, $A^\mu B_\mu = A \cdot B = A^0 B^0 - \bi{A} \cdot  \bi{B}$. The rest of the article is organized in the following manner. We start with introducing the formalism in Sec.~\ref{sec2} for our calculations. Then in Sec.~\ref{sec3}, we present the results along with some discussions. Finally, in Sec.~\ref{conclusion} we conclude our findings with an outlook to it.

\section{Formalism}
\label{sec2}
    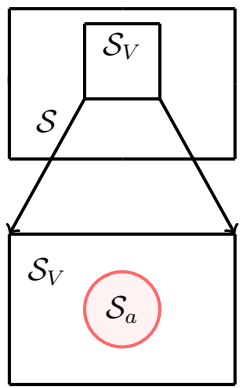
\begin{figure}[]
    \centering
\begin{tikzpicture}
\begin{scope}[very thick]
\draw [black](0,2) --(1.5,2);
\draw [black](1.5,2) --(3.,2);
\draw [black](0,0) --(0,1.1);
\draw [black](0,1.)--(0,2);
\draw [black](3,0) --(3,1.);
\draw [black](3,1.)--(3,2);
\draw [black](3,0) --(1.5,0);
\draw [black](1.5,0) --(0,0);
\draw [black](1.0,0.8) --(1.0,1.8);
\draw [black](1.0,0.8) --(2.0,0.8);
\draw [black](1.0,1.8) --(2.0,1.8);
\draw [black](2.0,1.8) --(2.0,0.8);
\draw [black](0,-1) --(0,-3);
\draw [black](0,-1) --(3,-1);
\draw [black](3,-1) --(3,-3);
\draw [black](0,-3) --(3,-3);
\draw [black,<-](0,-1) --(1.0,0.8);
\draw [black,<-](3,-1) --(2.0,0.8);\filldraw[color=red!60, fill=red!5, very thick](1.5,-2.0) circle (0.5);
\end{scope}
\node at (1.5,1.5) {$\mathcal{S}_V$};
\node at (0.5,0.5) {$\mathcal{S}$};
\node at (0.5,-1.5) {$\mathcal{S}_V$};
\node at (1.5,-2.0) {$\mathcal{S}_a$};
\end{tikzpicture}
\caption{Pictorial representation of the physical system: $\mathcal{S}$ is a closed/isolated system described by the microcanonical ensemble. $\mathcal{S}_V$ is a subsystem of the closed system in equilibrium, described by the canonical ensemble. $\mathcal{S}_a$ is a subsystem of $\mathcal{S}_V$ which is described by a ``Gaussian box". We calculate the quantum statistical fluctuation of energy within the subsystem $\mathcal{S}_a$~\cite{wojtek}. }
\label{fig1}
 \end{figure}
In the present investigation, we look into the quantum statistical fluctuation of energy within a subsystem $\mathcal{S}_a$ of the thermodynamic system $\mathcal{S}_V$. The thermodynamic system $\mathcal{S}_V$ is described by the canonical ensemble characterized by the temperature $T$ (or its inverse $\beta = 1/T$) (as depicted in Fig.~(\ref{fig1})). We assume that 
the characteristic volume of the sub-system $\mathcal{S}_a$ is less than the volume $V$ of the system $\mathcal{S}_V$. As mentioned earlier to calculate the quantum statistical fluctuation of energy we consider the time-time component of the energy-momentum tensor ($\hat{\mathcal{T}}^{tt}$) of the scalar field as well as the fermionic field. Following~\cite{Das:2021acta,Das:2021aar,Chen:2018cts}, we define a spatially smeared operator $\hat{\mathcal{T}}^{tt}_a$ that describes the time-time component of the energy-momentum tensor operator of a {\it finite} subsystem $\mathcal{S}_a$ placed at the origin of the coordinate system~\footnote{To remove any possible sharp boundary effects the Gaussian box has been introduced in Eq.~(\ref{equ1}) to define a smeared operator~\cite{Das:2021acta,Das:2021aar,Chen:2018cts,Feynman:100771}. Also note that the factor $(a\sqrt{\pi})^3$ in Eq.~(\ref{equ1}) has been introduced as normalization ~\cite{Chen:2018cts}.},
\begin{eqnarray}
\hat{\mathcal{T}}^{tt}_a = \frac{1}{(a\sqrt{\pi})^3}\int d^3 x~\hat{\mathcal{T}}^{tt}(x)~\exp\left(-\frac{{\bi{x}}^2}{a^2}\right).
\label{equ1}
\end{eqnarray}
In accordance with the general notion of quantum mechanics, we consider here the variance~\footnote{For a general quantum field theory operator $\hat{\mathcal{O}}$, a reasonable measure of the quantum spread in a given state can be defined by the variance of the operator $\hat{\mathcal{O}}$~\cite{Chen:2018cts}.}, 
\begin{eqnarray}
 \sigma^2(a,m,T) = \langle :\hat{\mathcal{T}}^{tt}_a: :\hat{\mathcal{T}}^{tt}_a: \rangle - \langle :\hat{\mathcal{T}}^{tt}_a :\rangle^2\, 
 \label{equ2}
\end{eqnarray}
and the normalized standard deviation,
\begin{eqnarray} 
\sigma_n(a,m,T)= \frac{(\langle:\hat{\mathcal{T}}^{tt}_a::\hat{\mathcal{T}}^{tt}_a:\rangle- \langle :\hat{\mathcal{T}}^{tt}_a :\rangle^2)^{1/2}}{\langle :\hat{\mathcal{T}}^{tt}_a :\rangle}.
\label{equ3}
\end{eqnarray}
as the measure of the fluctuation. In the definition of $\sigma^2$ and  $\sigma_n$ we have used the normal ordering of operators (denoted by the symbol `: :') to eliminate any zero-point energy contribution. Although, for a quantum field theory operator $\hat{\mathcal{O}}$ one can define the normal ordering prescription uniquely, i.e. $:\hat{\mathcal{O}:}$, but for a composite operator, e.g. $\hat{\mathcal{O}}\hat{\mathcal{O}}$ the normal ordering prescription to remove any divergent vacuum contribution is not unique. In the present work, we define the normal ordering of composite operator as    $:\hat{\mathcal{O}}::\hat{\mathcal{O}}:$. Alternately one may also define a normal ordered composite operator as $:\hat{\mathcal{O}}\hat{\mathcal{O}}:$~\cite{Kuo:1993if}, however, we argue in the subsequent sections that our definition of the normal ordering, i.e. $:\hat{\mathcal{O}}::\hat{\mathcal{O}}:$ reproduces correctly limiting results for the energy fluctuation. Furthermore in Eqs.~(\ref{equ2}) and (\ref{equ3}), ``$\langle \hat{\mathcal{O}}\rangle$" denotes the thermal average/thermal expectation of an operator $\hat{\mathcal{O}}$. For the estimation of quantum statistical fluctuations, we evaluate $\sigma^2$ and $\sigma_n$ for a thermal system consisting of spin-zero as well as spin-half particles. For this purpose, we need to find an appropriate expression of $\hat{\mathcal{T}}^{tt}_a $ along with the prescription of thermal averaging.

\subsection{Scalar field}
To describe a system by a quantum scalar field in thermal equilibrium we consider the Fourier space representation of a real scalar field~\cite{Das:2021acta,Chen:2018cts}
\begin{eqnarray}
\phi(t,{\bi{x}})=\int\frac{d^3p}{\sqrt{(2\pi)^3 ~2\omega_{\bi{p}}}}\left(\mathfrak{a}_{\bi{p}}^{\pdagger}e^{-i p \cdot x} +
\mathfrak{a}_{\bi{p}}^{\dagger}e^{i p \cdot x} \right),
\label{equ4}
\end{eqnarray}
here $\mathfrak{a}_{\bi{p}}^{\pdagger}$  is the bosonic annihilation operator and  $\mathfrak{a}_{\bi{p}}^{\dagger}$ is the bosonic creation operator  satisfying the standard canonical commutation relations $[\mathfrak{a}_{\bi{p}}^{\pdagger},\mathfrak{a}_{\bi{p}^{\prime}}^{\dagger}] = \delta^{(3)}(\bi{p}-\bi{p}^{\prime})$.  $\omega_{\bi{p}}=\sqrt{{\bi{p}}^2+m^2}$ is the single particle dispersion relation.
For the scalar field as defined in Eq.~(\ref{equ4}) the time-time component of the energy-momentum tensor operator is noting but the canonical Hamiltonian density, i.e., $\hat{\mathcal{T}}^{tt}_{\phi}\equiv \mathcal{H}_{\phi}= (\dot{\phi}^2+({\bf{\nabla}\phi)}^2+m^2\phi^2)/2$. Here $\dot{\phi}$ and $\bf{\nabla}\phi$ represents time and space derivative of the scalar field, respectively. For the scalar field to calculate the variance $(\sigma_{\phi}^2)$ and the normalized standard deviation $(\sigma_{n,\phi})$ we have to perform the thermal averaging involving the smeared operator $\hat{\mathcal{T}}^{tt}_{a,\phi}$ or $\mathcal{H}_{a,\phi}$. For such thermal averaging
it is sufficient to know the thermal expectation values of the products of two and four creation and/or annihilation operators~\cite{Das:2021acta,CohenTannoudji:422962,Itzykson:1980rh,Evans:1996bha}, i.e.,
\begin{eqnarray}
 & \langle \mathfrak{a}^{\dagger}_{{\bi{p}}}\mathfrak{a}_{{\bi{p}}^{\prime}}^{\pdagger}\rangle=\delta^{(3)}({\bi{p}}-{\bi{p}}^{\prime})\mathfrak{f}_b(\omega_{{\bi{p}}}),\label{equ5}\\
& \langle \mathfrak{a}^{\dagger}_{{\bi{p}}}\mathfrak{a}^{\dagger}_{{\bi{p}}^{\prime}}\mathfrak{a}_{{\bi{k}}}^{\pdagger}\mathfrak{a}_{{\bi{k}}^{\prime}}^{\pdagger}\rangle = \bigg(\delta^{(3)}({\bi{p}}-{\bi{k}})~\delta^{(3)}({\bi{p}}^{\prime}-{\bi{k}}^{\prime})\nonumber\\
& ~~~~~~~~~~~~~~~~~~~~~~~+\delta^{(3)}({\bi{p}}-{\bi{k^{\prime}}})~\delta^{(3)}({\bi{p}}^{\prime}-{\bi{k}})\bigg)\mathfrak{f}_b(\omega_{{\bi{p}}})\mathfrak{f}_b(\omega_{{\bi{p}}^{\prime}}). \label{equ6}
\end{eqnarray}
In these equations, $\mathfrak{f}_b(\omega_{{\bi{p}}})=1/(\exp[\beta ~\omega_{{\bi{p}}}]-1)$ is the Bose--Einstein distribution function. Any other combinations of two and four creation and/or annihilation operators which may appear in the calculation of the thermal averaging of composite operators can be obtained from Eqs.~(\ref{equ5}) and (\ref{equ6})
using the commutation relation between $\mathfrak{a}_{{\bi{p}}}^{\pdagger}$ and $\mathfrak{a}_{{\bi{p}}}^{\dagger}$.

\subsection{Dirac field}
To describe the system of spin-half particles in thermal equilibrium we consider the Dirac field operator~\cite{Das:2021aar,Tinti:2020gyh}
\begin{eqnarray}
\psi(t,\bi{x})=\sum_r\int\frac{d^3p}{(2\pi)^3\sqrt{2\omega_{\bi{ p}}}}\Big(\mathcal{U}_r^{\pdagger}(\bi{p})a_r^{\pdagger}(\bi{p})e^{-i p \cdot x}+\mathcal{V}_r^{\pdagger}(\bi{p})b_r^{\dagger}(\bi{p})e^{i p \cdot x} \Big),
\label{equ7}
\end{eqnarray}
here $a_r^{\pdagger}(\bi{p})$ is particle annihilation operator and $b_r^{\dagger}(\bi{p})$ is the antiparticle creation operator satisfying anti-commutation relations, $\{a_r^{\pdagger}(\bi{p}),a_s^{\dagger}(\bi{p}^{\prime})\} =(2\pi)^3\delta_{rs} \delta^{(3)}(\bi{p}-\bi{p}^{\prime})$ and
$ \{b_r^{\pdagger}(\bi{p}),b_s^{\dagger}(\bi{p}^{\prime})\} =(2\pi)^3\delta_{rs} \delta^{(3)}(\bi{p}-\bi{p}^{\prime})$. The index $r$ denotes the polarization degree of freedom. The Dirac spinors $\mathcal{U}_r^{\pdagger}(\bi{p})$ and $\mathcal{V}_r^{\pdagger}(\bi{p})$ are normalized as, $\bar{\mathcal{U}}_r^{\pdagger}(\bi{p}) \mathcal{U}_s^{\pdagger}(\bi{p}) = 2 m \delta_{rs}$ and $\bar{ \mathcal{V}}_r^{\pdagger}(\bi{p}) \mathcal{V}_s^{\pdagger}(\bi{p}) = -2 m \delta_{rs}$. $\omega_{\bi{p}}=\sqrt{\bi{p}^2+m^2}$ is again the single particle energy.

Contrary to the energy-momentum tensor of a real scalar field, the canonical energy-momentum tensor of spin half particles as obtained using the Noether theorem is not symmetric. One can use a pseudo-gauge transformation to obtain a symmetric energy-momentum tensor~\cite{HEHL197655,Speranza:2020ilk,Florkowski:2018fap}. We should emphasize that the symmetric form of the energy-momentum tensor naturally appears in the general theory of relativity, where one obtains the energy-momentum tensor using the variation of the action with respect to the space-time metric. In literature various forms of symmetric energy-momentum tensor have been considered, e.g., Belinfante-Rosenfeld energy-momentum tensor~\cite{BELINFANTE1939887,BELINFANTE1940449,Rosenfeld1940}, etc. To study any effect of the form of various energy-momentum tensor or pseudo-gauge transformation, we consider the canonical form of the energy-momentum tensor along with the Belinfante-Rosenfeld energy-momentum tensor~\cite{BELINFANTE1939887,BELINFANTE1940449,Rosenfeld1940,Tinti:2020gyh}, de~Groot-van~Leeuwen-van~Weert energy-momentum tensor~\cite{DeGroot:1980dk} and Hilgevoord-Wouthuysen energy-momentum tensor~\cite{HILGEVOORD19631,HILGEVOORD19651002} to obtain quantum statistical fluctuation of energy.  

\subsubsection{Canonical EMT:} The canonical form of the energy-momentum tensor of a spin half field can be shown to be~\cite{Das:2021aar,Tinti:2020gyh},

 \begin{equation}
     \hat{\mathcal{T}}^{\mu\nu}_{\psi,Can}=\frac{i}{2}\bar\psi\gamma^{\mu}\mathcal{D}^{\nu}\psi,
     \label{equ8}
 \end{equation}
where $\mathcal{D}^{\mu}\equiv\overrightarrow{\partial}^{\mu}-\overleftarrow{\partial}^{\mu}$. In order to obtain the canonical energy-momentum tensor as given in Eq.~(\ref{equ8}) one uses the Dirac equation for $\psi$ and $\bar{\psi}$~\cite{Tinti:2020gyh}. Note the canonical energy momentum tensor is not manifestly symmetric under $\mu\leftrightarrow \nu$ exchange. Using the ``$tt$" component of the canonical energy-momentum tensor we can obtain $\sigma^2_{\psi,Can}$ and $\sigma_{n,\psi,Can}$ of the smeared operator $\hat{\mathcal{T}}^{tt}_{\psi,Can,a}$ for the subsystem $\mathcal{S}_a$.

\subsubsection{Belinfante-Rosenfeld EMT:}
The Belinfante-Rosenfeld improved symmetric energy-momentum tensor has the following form~\cite{Das:2021aar,Tinti:2020gyh}
\begin{eqnarray}
    \hat{\mathcal{T}}^{\mu\nu}_{\psi,BR}=\frac{i}{2}\bar{\psi}\gamma^{\mu}\mathcal{D}^\nu\psi-\frac{i}{16}\partial_{\lambda}\Big(\bar{\psi}\Big\{\gamma^{\lambda},\Big[\gamma^{\mu},\gamma^{\nu}\Big]\Big\}\psi\Big).
\label{equ9}
\end{eqnarray}
Note that the Belinfante-Rosenfeld improved energy-momentum tensor as given in Eq.~(\ref{equ9}) is also not manifestly symmetric under $\mu\leftrightarrow \nu$ exchange. However using the equation of motion for the Dirac fields one can argue that $\hat{\mathcal{T}}^{\mu\nu}_{\psi,BR}$ is symmetric under $\mu\leftrightarrow \nu$ exchange~\cite{Tinti:2020gyh}. From Eqs.~(\ref{equ8}) and (\ref{equ9})  it is clear that $\hat{\mathcal{T}}^{tt}_{\psi,BR} = \hat{\mathcal{T}}^{tt}_{\psi,Can}$. Therefore, thermal average of the normal ordered $\hat{\mathcal{T}}^{tt}_{\psi,BR,a}$ and its fluctuation will be the same as obtained in the canonical framework. 

\subsubsection{de~Groot-van~Leeuwen-van~Weert EMT:}
Another very important form of the symmetric energy-momentum tensor is the de~Groot-van~Leeuwen-van~Weert energy-momentum tensor for a massive spin-half field. It can be expressed as~\cite{Das:2021aar,DeGroot:1980dk,Tinti:2020gyh},
\begin{eqnarray}
    \hat{\mathcal{T}}^{\mu\nu}_{\psi,GLW}&= -\frac{1}{4m}\bar{\psi}\mathcal{D}^{\mu}\mathcal{D}^{\nu}\psi-g^{\mu\nu}\mathcal{L}_{D}\nonumber\\
    & =\frac{1}{4m}\Big[-\bar{\psi}(\partial^{\mu}\partial^{\nu}\psi)+(\partial^{\mu}\bar{\psi})(\partial^{\nu}\psi) +(\partial^{\nu}\bar{\psi})(\partial^{\mu}\psi)\nonumber\\
    &~~~~~~~~~~~-(\partial^{\mu}\partial^{\nu}\bar{\psi})\psi\Big].
    \label{equ10}
\end{eqnarray}
Here $\mathcal{L}_{D}\equiv\frac{i}{2}\bar\psi\gamma^{\mu}\mathcal{D}_{\mu}\psi-m\bar{\psi}\psi$ is the Lagrangian density of a spin half field. The GLW symmetric energy-momentum tensor is manifestly symmetric under $\mu\leftrightarrow\nu$ exchange. Using the ``$tt$" component of the GLW energy-momentum tensor we can obtain the fluctuation, i.e. $\sigma^2_{\psi,GLW}$ and $\sigma_{n,\psi,GLW}$ of the smeared operator $\hat{\mathcal{T}}^{tt}_{\psi,GLW,a}$ for the subsystem $\mathcal{S}_a$.

\subsubsection{Hilgevoord-Wouthuysen EMT:}
Finally we also consider the Hilgevoord-Wouthuysen form of the  symmetric energy-momentum tensor~\cite{Das:2021aar,HILGEVOORD19631,HILGEVOORD19651002,Tinti:2020gyh},
\begin{eqnarray}
\hat{\mathcal{T}}^{\mu\nu}_{\psi,HW}&= \hat{\mathcal{T}}^{\mu\nu}_{\psi,Can} +\frac{i}{2m} \bigg(\partial^{\nu}\psbar \sigma^{\mu\beta}\partial_\beta \psi +\partial_\alpha \psbar \sigma^{\alpha\mu}\partial^{\nu}\psi\bigg) \nonumber\\
&- \frac{i}{4m}g^{\mu\nu} \partial_\lambda \left(\psbar\sigma^{\lambda\alpha}\mathcal{D}_\alpha \psi\right),
\end{eqnarray}
here $\sigma^{\mu\nu} \equiv (i/2) \left[\gamma^\mu,\gamma^\nu \right]$. Here also, using the ``$tt$" component of the HW energy-momentum tensor we can obtain the fluctuation, i.e. $\sigma^2_{\psi,HW}$ and $\sigma_{n,\psi,HW}$ of the smeared operator $\hat{\mathcal{T}}^{tt}_{\psi,HW,a}$ for the subsystem $\mathcal{S}_a$.

%Note that $\hat{\mathcal{T}}^{\mu\nu}_{\psi,Can}$ is not symmetric under $\mu\leftrightarrow\nu$ exchange. The anti-symmetric part of $\hat{\mathcal{T}}^{\mu\nu}_{\psi,Can}$ cancels the anti-symmetric part arising from other terms giving rise to a  symmetric energy momentum tensor $\hat{\mathcal{T}}^{\mu\nu}_{\psi,HW}$. 

Once we have defined different forms of the energy-momentum tensor of a spin-half field now we discuss the thermal averaging prescription. Similar to the scalar field here also to perform thermal averaging we only need to know the thermal expectation values of the products of two and four creation and/or annihilation operators, but for both particles and antiparticles ~\cite{Das:2021aar,CohenTannoudji:422962,Itzykson:1980rh,Evans:1996bha}. The thermal average of the products of two and four creation and/or annihilation operators can be expressed as~\cite{Das:2021aar,CohenTannoudji:422962,Itzykson:1980rh,Evans:1996bha},
\begin{eqnarray}
& \langle a_r^{\dagger}({\bi{p}})a_s^{\pdagger}({\bi{p}}^{\prime})\rangle=(2\pi)^3\delta_{rs}\delta^{(3)}({\bi{p}}-{\bi{p}}^{\prime})\mathfrak{f}_f(\omega_{\bi{p}}),\label{equ12}\\
& \langle a^{\dagger}_r(\bi{p})a^{\dagger}_s(\bi{p}^{\prime})a_{r^{\prime}}^{\pdagger}(\bi{k})a_{s^{\prime}}^{\pdagger}(\bi{k}^{\prime})\rangle\nonumber\\
& =(2\pi)^6 \Big(\delta_{rs^{\prime}}\delta_{r^{\prime}s}\delta^{(3)}(\bi{p}-\bi{k}^{\prime})~\delta^{(3)}(\bi{p}^{\prime}-\bi{k})\nonumber\\
&-\delta_{rr^{\prime}}\delta_{ss^{\prime}}\delta^{(3)}({\bi{p}}-\bi{k})~\delta^{(3)}({\bi{p}}^{\prime}-\bi{k}^{\prime})\Big)\mathfrak{f}_f(\omega_{{\bi{p}}})\mathfrak{f}_f(\omega_{{\bi{p}}^{\prime}}).\label{equ13}
\end{eqnarray}
In these equations $\mathfrak{f}_f(\omega_{{\bi{p}}})$ is the Fermi--Dirac distribution function for particles. The operators associated with anti-particles also satisfies similar equations as given in Eqs.~(\ref{equ12}) and (\ref{equ13}). But for antiparticles, the corresponding  Fermi--Dirac distribution function differs by the sign of the chemical potential associated with the conserved charge~\cite{Das:2021rck}. In the present case, we do not consider any chemical potential.

\section{Results and discussions}
\label{sec3}
Once we have defined all the relevant operators and the rules for the thermal averaging, now we present the analytical as well as the numerical estimation of quantum statistical fluctuation of $\hat{\mathcal{T}}^{tt}_a$ for the bosonic and fermionic system. We present our results to emphasize three different aspects of quantum statistical fluctuations. These are, (i) scaling or variation of fluctuation with the system size $(a)$, temperature ($T$) and particle mass ($m$), (ii) dependence of quantum statistical fluctuation on the form of the energy-momentum tensor operator or the pseudo-gauge dependence of quantum statistical fluctuations and (iii) we also argue that in the large volume limit the quantum statistical fluctuation obtained in a subsystem gives rise to the statistical fluctuation already known from statistical physics.  
\subsection{Analytical formula}
Before we go into any numerical analysis we first present analytical results.
\subsubsection{For spin zero field:}
As argued above using the expression of the canonical Hamiltonian density of a real scalar field one can obtain the smeared operator  $ \mathcal{H}_{a,\phi}$  following the definition as given in Eq.~(\ref{equ1}). Using $ \mathcal{H}_{a,\phi}$ along with the normal ordering and thermal averaging prescriptions one can get the thermal expectation value of $: \mathcal{H}_{a,\phi}:$ and the corresponding fluctuation $\sigma_{\phi}^2$ or $\sigma_{n,\phi}$.  The thermal expectation value of the operator $ \mathcal{H}_{a,\phi}$ can be expressed as~\cite{Das:2021acta}, 
\begin{equation}
    \langle :\mathcal{H}_{a,\phi} :\rangle= \int \frac{d^3{{p}}}{(2\pi)^3}~\omega_{{\bi{p}}}~\mathfrak{f}_b\left(\omega_{{\bi{p}}}\right) \equiv \varepsilon(T,m),
    \label{equ14}
\end{equation}
which is energy density. This result is consistent with the result known from the kinetic-theory considerations~\cite{Huang:1987asp}. It is interesting to note that due to the spatial uniformity of the system  $\langle :\mathcal{H}_{a,\phi} :\rangle$ is independent of scale $a$. Furthermore these results are also time independent. Although $\langle :\mathcal{H}_{a,\phi} :\rangle$ is system size independent but we show that the quantum fluctuation, i.e. $\sigma_{\phi}^2$ as defined in Eq.~(\ref{equ2}) explicitly depends upon the scale `$a$'. Using the rules for the thermal expectation values as given in Eqs.~(\ref{equ5})-(\ref{equ6}) and dropping all the temperature independent vacuum energy terms,  it can be shown that~\cite{Das:2021acta} 
\begin{eqnarray}
\sigma^2_{\phi}(a,m,T) &=  \int dP ~dP^{\prime} \mathfrak{f}_b(\omega_{{\bi{p}}})(1+\mathfrak{f}_b(\omega_{{\bi{p}}^{\prime}}))\nonumber\\
&\times \bigg[(\omega_{{\bi{p}}}\omega_{{\bi{p}}^{\prime}}+{\bi{p}}\cdot{\bi{p}}^{\prime}+m^2)^2e^{-\frac{a^2}{2}({\bi{p}}-{\bi{p}}^{\prime})^2}\nonumber\\
&+(\omega_{{\bi{p}}}\omega_{{\bi{p}}^{\prime}}+{\bi{p}}\cdot{\bi{p}}^{\prime}-m^2)^2e^{-\frac{a^2}{2}({\bi{p}}+{\bi{p}}^{\prime})^2}\bigg],
\label{equ15}
\end{eqnarray}
where $dP = d^3{{p}}/((2\pi)^{3} 2 \omega_{{\bi{p}}})$. Eq.~(\ref{equ15}) represents the result for the energy fluctuations in ``Gaussian'' subsystem $\mathcal{S}_a$ of the system $\mathcal{S}_V$. This formula also allows us to determine variation of quantum statistical fluctuation for subsystem size ($a$), temperature ($T$), and particle mass ($m$). The integrals as given in Eq.~(\ref{equ15}) can be done analytically for a massless particle with Boltzmann statistics to obtain an analytical formula for $\sigma_{\phi}^2$~\cite{Das:2021acta}. We should note that the particles can  have internal degrees of freedom, e.g. isospin, color charge, etc. These internal degrees of freedom can be included by introducing the degeneracy factor ($g_b$). Technically this can be achieved by replacements $\varepsilon \rightarrow g_b \varepsilon$ and $\sigma_{\phi}^2 \rightarrow g_b \sigma_{\phi}^2$~\cite{Das:2021acta}.

\subsubsection{For spin-half field:} Analogous to the scalar field, for the spin-half field also, we obtain the smeared operator $:\hat{\mathcal{T}}^{tt}_{\psi,a}:$. However, we should note that for the fermionic system we have different choices of the energy-momentum tensor. Using the prescription of the thermal average for the spin half field we find~\cite{Das:2021aar}, 
\begin{eqnarray}
\langle:\hat{\mathcal{T}}^{tt}_{\psi,Can,a}:\rangle
& = 4\int\frac{d^3p}{(2\pi)^3}~\omega_{\bi{p}}~\mathfrak{f}_f(\omega_{\bi{p}}) \equiv\varepsilon_{Can}(T,m) \nonumber\\
& =\langle:\hat{\mathcal{T}}^{tt}_{\psi,BR,a}:\rangle=\langle:\hat{\mathcal{T}}^{tt}_{\psi,GLW,a}:\rangle=\langle:\hat{\mathcal{T}}^{tt}_{\psi,HW,a}:\rangle.
\label{equ16}
\end{eqnarray}
Although the exact expressions of energy-momentum tensor is pseudo-gauge dependent but the thermal expectation value of $:\hat{\mathcal{T}}^{tt}_{\psi,a}:$ does not depend on pseudo-gauge, i.e., $\varepsilon_{Can}(T,m)=\varepsilon_{BR}(T,m)=\varepsilon_{GLW}(T,m)=\varepsilon_{HW}(T,m)$. One should note that energy density for spin-half field as obtained here is also independent of the scale `$a$'. Now let us look into the expression of the quantum fluctuation for different forms of the energy-momentum tensor~\cite{Das:2021aar}, 
\begin{eqnarray}
&\sigma^2_{\psi,Can}(a,m,T) =  2\int dP ~dP^{\prime} \mathfrak{f}_f(\omega_{{\boldsymbol{p}}})(1-\mathfrak{f}_f(\omega_{{\boldsymbol{p}}^{\prime}}))\nonumber\\
&\times \bigg[(\omega_{{\boldsymbol{p}}}+\omega_{{\boldsymbol{p}}^{\prime}})^2(\omega_{{\boldsymbol{p}}}\omega_{{\boldsymbol{p}}^{\prime}}+{\boldsymbol{p}}\cdot{\boldsymbol{p}}^{\prime}+m^2)e^{-\frac{a^2}{2}({\boldsymbol{p}}-{\boldsymbol{p}}^{\prime})^2}\nonumber\\
&-(\omega_{{\boldsymbol{p}}}-\omega_{{\boldsymbol{p}}^{\prime}})^2(\omega_{{\boldsymbol{p}}}\omega_{{\boldsymbol{p}}^{\prime}}+{\boldsymbol{p}}\cdot{\boldsymbol{p}}^{\prime}-m^2)e^{-\frac{a^2}{2}({\boldsymbol{p}}+{\boldsymbol{p}}^{\prime})^2}\bigg],
\label{equ17}
\end{eqnarray}

\begin{eqnarray}
&\sigma^2_{\psi,GLW}(a,m,T) =  \frac{1}{2m^2}\int dP ~dP^{\prime} \mathfrak{f}_f(\omega_{{\boldsymbol{p}}})(1-\mathfrak{f}_f(\omega_{{\boldsymbol{p}}^{\prime}}))\nonumber\\
&\times \bigg[(\omega_{{\boldsymbol{p}}}+\omega_{{\boldsymbol{p}}^{\prime}})^4\left(\omega_{{\boldsymbol{p}}}\omega_{{\boldsymbol{p}}^{\prime}}-{\boldsymbol{p}}\cdot{\boldsymbol{p}}^{\prime}+m^2\right)e^{-\frac{a^2}{2}({\boldsymbol{p}}-{\boldsymbol{p}}^{\prime})^2}\nonumber\\
&-(\omega_{{\boldsymbol{p}}}-\omega_{{\boldsymbol{p}}^{\prime}})^4 \left(\omega_{{\boldsymbol{p}}}\omega_{{\boldsymbol{p}}^{\prime}}-{\boldsymbol{p}}\cdot{\boldsymbol{p}}^{\prime}-m^2\right)e^{-\frac{a^2}{2}({\boldsymbol{p}}+{\boldsymbol{p}}^{\prime})^2}\bigg],
\label{equ18}
\end{eqnarray}
and, 
\begin{eqnarray}
&\sigma^2_{\psi,HW}(a,m,T) =  \frac{2}{m^2}\int dP ~dP^{\prime} \mathfrak{f}_f(\omega_{{\boldsymbol{p}}})(1-\mathfrak{f}_f(\omega_{{\boldsymbol{p}}^{\prime}}))\nonumber\\
&\times \Big[\left(\omega_{{\boldsymbol{p}}}\omega_{{\boldsymbol{p}}^{\prime}}+{\boldsymbol{p}}\cdot{\boldsymbol{p}}^{\prime}+m^2\right)^2\left(\omega_{{\boldsymbol{p}}}\omega_{{\boldsymbol{p}}^{\prime}}-{\boldsymbol{p}}\cdot{\boldsymbol{p}}^{\prime}+m^2\right) e^{-\frac{a^2}{2}({\boldsymbol{p}}-{\boldsymbol{p}}^{\prime})^2}\nonumber\\
&~~-(\omega_{{\boldsymbol{p}}}\omega_{{\boldsymbol{p}}^{\prime}}+{\boldsymbol{p}}\cdot{\boldsymbol{p}}^{\prime}-m^2)^2(\omega_{{\boldsymbol{p}}}\omega_{{\boldsymbol{p}}^{\prime}}-{\boldsymbol{p}}\cdot{\boldsymbol{p}}^{\prime}-m^2)e^{-\frac{a^2}{2}({\boldsymbol{p}}+{\boldsymbol{p}}^{\prime})^2}\Big],
\label{equ19}
\end{eqnarray}
here $dP \equiv d^3{{p}}/((2\pi)^{3} 2 \omega_{{\bi{p}}})$.
From Eqs.~(\ref{equ17})-(\ref{equ19}) one can observe that the quantum fluctuation is indeed depends on the form of the energy-momentum tensor or the pseudo-gauge choice. Here also the degeneracy factor can be incorporated in the following manner:
$\varepsilon_{Can} \rightarrow g_f \varepsilon_{Can}$ and $\sigma_{\psi}^2 \rightarrow g_f \sigma_{\psi}^2$.

\subsubsection{Large volume limit or the thermodynamic limit:}
Since $\mathcal{S}_a$ is a subsystem of the larger system $\mathcal{S}_V$, we expect that in the large volume limit $a\rightarrow \infty$, but still keeping $a^3 \ll V$,  formula for quantum statistical fluctuation should reproduce the known result from statistical mechanics~\cite{Huang:1987asp}. This can be checked using the Gaussian representation of the three dimensional Dirac delta function, i.e. 
\begin{equation}
    \delta^{(3)}({\bi{p}}-{\bi{p}^{\prime}})=\lim_{a \to\infty} \frac{a^3}{(2\pi)^{3/2}}e^{-\frac{a^2}{2}({\bi{p}}-{\bi{p}^{\prime}})^2}.
    \label{equ20}
\end{equation}
In the large volume limit Eq.~(\ref{equ15}) boils down to~\cite{Das:2021acta},  
\begin{equation}
\sigma_{\phi}^2 = \frac{g_b}{(2\pi)^{3/2} a^3}
\int \frac{d^3{{p}}}{(2\pi)^3}~\omega_{{\bi{p}}}^2~\mathfrak{f}_b(\omega_{{\bi{p}}}) (1+\mathfrak{f}_b(\omega_{{\bi{p}}})).
\label{equ21}
\end{equation}
The factor $\mathfrak{f}(\omega_{{\bi{p}}}) (1+\mathfrak{f}(\omega_{{\bi{p}}}))$ in the above equation gives us the hint that $\sigma_{\phi}^2$ can be expressed in term of the specific heat $C_V$. This in indeed true. Note that using the expression of the energy density as given in Eq.~(\ref{equ14}) $C_V$ can be expressed as, 
\begin{equation} 
C_{V,\phi} = \frac{d\varepsilon}{dT} = \frac{g_b}{T^2} \int \frac{d^3{{p}}}{(2\pi)^3}~\omega_{{\bi{p}}}^2~\mathfrak{f}_b(\omega_{{\bi{p}}}) (1+\mathfrak{f}_b(\omega_{{\bi{p}}})).
\label{equ22}
\end{equation}
Using Eqs.~(\ref{equ21})-(\ref{equ22}) we can conclude that in the large volume limit~\cite{Das:2021acta,Mrowczynski:1997mj}, 
\begin{equation}
V_a \sigma_{n,\phi}^2
= \frac{T^2 C_{V,\phi}}{\varepsilon^2}
= V \frac{\langle E^2\rangle-\langle E \rangle^2}{\langle E \rangle^2} \equiv V \sigma^2_E,
\label{equ23}
\end{equation}
here $V_a = a^3 (2\pi)^{3/2}$ may be considered as the volume of the ``Gaussian'' subsystem $\mathcal{S}_a$~\cite{Das:2021acta,Das:2021aar}. Furthermore in the above equation $E$ represents the energy of system $\mathcal{S}_V$ and $V\sigma_E^2$ denotes the normalized energy fluctuation in the system $\mathcal{S}_V$~\cite{Huang:1987asp}. Therefore we observe that in the large volume limit the volume scaled quantum statistical fluctuation $V_a\sigma_{n,\phi}^2$ reproduces the correct thermodynamic fluctuation $V\sigma^2_E$ \footnote{Since we could reproduce the correct fluctuations in the large volume limit, this also shows that the choice of the normal ordering prescription as considered here is physically more appealing to obtain quantum statistical fluctuation.}. Such conclusion can also be obtained for a fermionic system. However for spin-half particles something more interesting happens. Earlier we have shown that for spin-half field quantum statistical fluctuation depends on the form of the energy-momentum tensor, i.e. generically, $\sigma^2_{\psi,Can}\neq\sigma^2_{\psi,GLW}\neq\sigma^2_{\psi,HW}$. However if we consider the large volume limit then it can be argued that~\cite{Das:2021aar}, 
\begin{equation}
\sigma_{\psi,Can}^2 = ~ 
\frac{4~g_f}{(2\pi)^{3/2} a^3}
\int \frac{d^3{{p}}}{(2\pi)^3}~\omega_{{\bi{p}}}^2~\mathfrak{f}_f(\omega_{{\bi{p}}}) (1-\mathfrak{f}_f(\omega_{{\bi{p}}})) = \sigma_{\psi,GLW}^2 = \sigma_{\psi,HW}^2 . 
\label{equ24}
\end{equation}
Hence in the large volume limit the energy-fluctuation does not depend on the pseudo-gauge choice of the energy-momentum tensor. More explicitly\footnote{Here we do not introduce different notations to represent statistical fluctuation of energy for the system $\mathcal{S}_V$ to distinguish between bosonic and fermionic systems. Rather, $V \sigma^2_E$ represents statistical fluctuation within the system $\mathcal{S}_V$ as described by the canonical ensemble for both spin-zero and spin half particles. Difference between the expression of $V \sigma^2_E$ as given in Eq.~(\ref{equ23}) and Eq.~(\ref{equ25}) is implicitly understood.}, 
\begin{equation}
V_a \sigma_{n,\psi,Can}^2=V_a \sigma_{n,\psi,GLW}^2
= \frac{T^2 C_{V,\psi}}{\varepsilon_{Can}^2}
= V \frac{\langle E^2\rangle-\langle E \rangle^2}{\langle E \rangle^2} \equiv V \sigma^2_E.
\label{equ25}
\end{equation}

\subsection{Numerical results}
\begin{figure}
    \centering
    \begin{minipage}{.48\textwidth}
        \centering
        \includegraphics[width=1.2\linewidth]{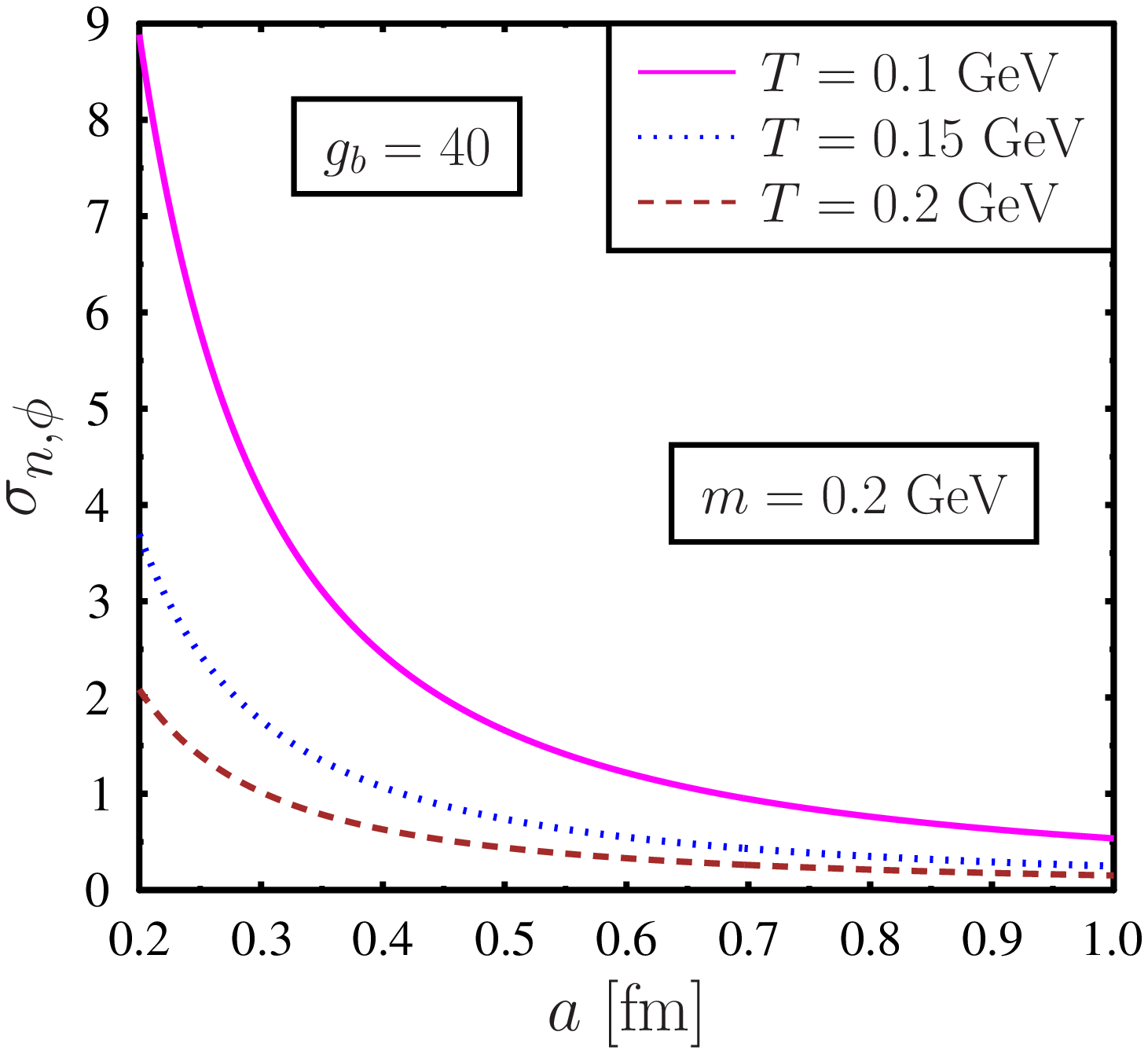}
    \end{minipage}~~
    \begin{minipage}{0.48\textwidth}
        \centering
        \includegraphics[width=1.2\linewidth]{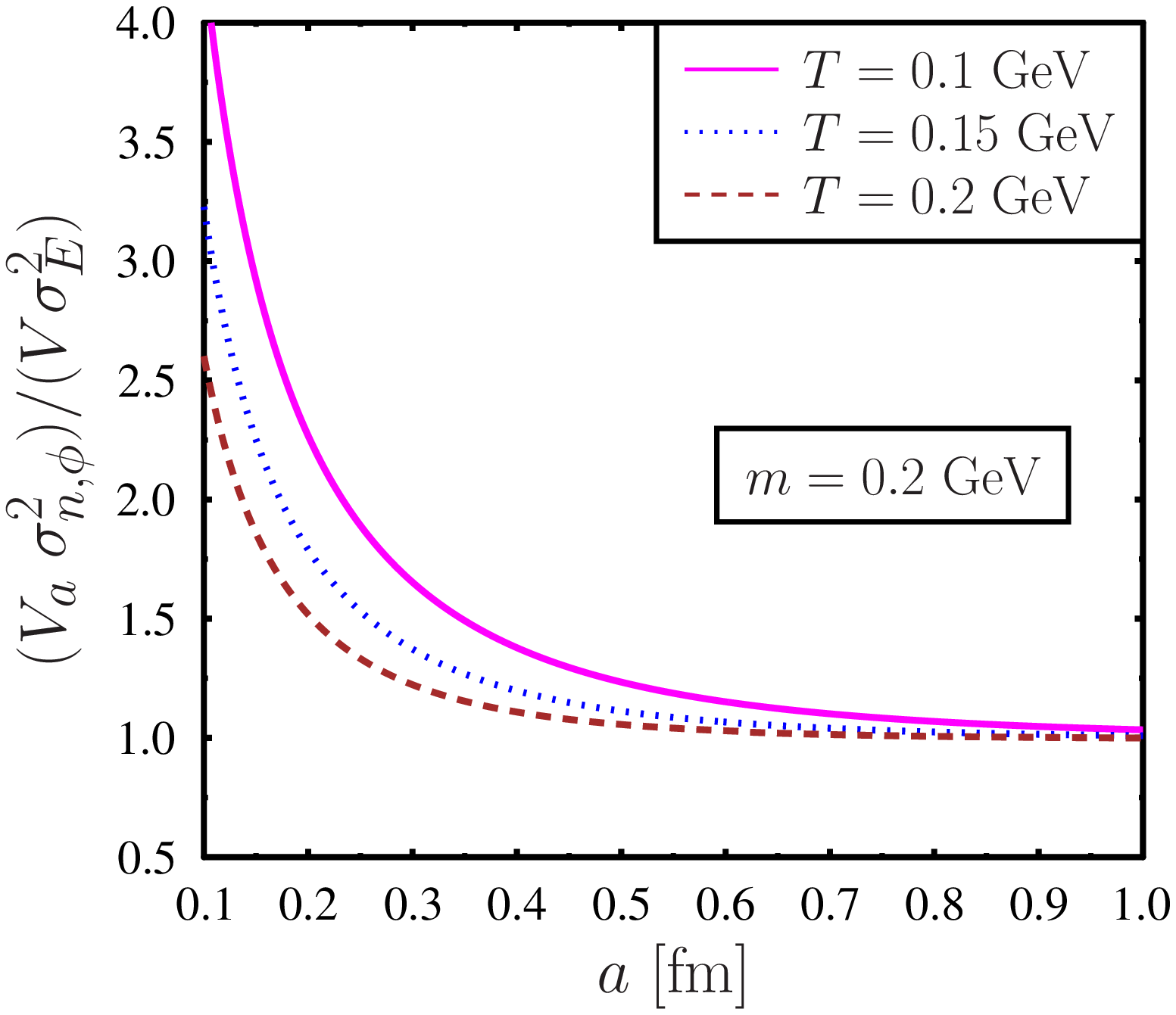}
    \end{minipage}
    \caption{Left plot: variation of normalized fluctuation $\sigma_{n,\phi}$ within the subsystem $\mathcal{S}_a$ described by a real scalar field with the scale `$a$' for different temperature $(T)$ but fixed mass $(m)$.  Right plot: in this plot we demonstrate the large system size limit by plotting the normalized quantity $V_a \sigma_{n,\phi}^2/V \sigma^2_E$ with the scale `$a$' for different temperature $(T)$ but fixed mass $(m)$. For more discussions see text and Ref.~\cite{Das:2021acta}.}
\label{fig2}
\end{figure}

Once we have discussed the analytical results in the previous sections, now we present numerical results. In the left plot of Fig.~(\ref{fig2}) we show the variation of normalized fluctuation $\sigma_{n,\phi}$ within the subsystem $\mathcal{S}_a$ described by a real scalar field with the scale `$a$' for different temperature $(T)$ but fixed mass $(m)$~\footnote{While presenting numerical results we keep in mind the strongly interacting matter produced in heavy-ion collision experiments. For the numerical estimation of fluctuation we consider $100$~MeV $< T < 400$~MeV, and particle masses: $0$~MeV $< m < 1000$~MeV. We consider the total degeneracy factor to be $40$, which is a typical degeneracy factor for three flavor QCD matter. For a detailed discussion on the degeneracy factor see Ref.~\cite{Das:2021acta}. Since we are taking the total degeneracy factor to be $40$, we consider $g_b=40$ and $g_f=10$ for numerical results. This is because contrary to $\sigma_{\phi}^2$ or $\sigma_{n,\phi}$, $\sigma_{\psi}^2$ or $\sigma_{n,\psi}$ already includes the spin degeneracy factor and particle-antiparticle degeneracy factor.}. From this plot one may observe that with temperature normalized fluctuation $\sigma_{n,\phi}$ decreases. Note that although $\sigma_{n,\phi}$ decreases with temperature, but the variation of $\sigma_{n,\phi}$ with temperature is opposite to the variation of $\sigma_{\phi}$ ( not shown here explicitly). Furthermore with the system size `$a$' quantum fluctuation decreases and for $a\rightarrow 0$ quantum fluctuation diverges. This is nothing but the manifestation of quantum uncertainty principle. In the right plot of Fig.~(\ref{fig2}) we show the variation of  the normalized quantity $V_a \sigma_{n,\phi}^2/V \sigma^2_E$ with the scale `$a$' for different temperature $(T)$ but fixed mass $(m)$. As argued previously that in the large volume limit or in the thermodynamic limit $V_a \sigma_{n,\phi}^2$ should approach $V \sigma^2_E$, i.e. $V_a \sigma_{n,\phi}^2/V \sigma^2_E$ should approach unity in the thermodynamic limit. From this plot one may observe that for the range of temperature and mass, for $a>1$ fm $V_a \sigma_{n,\phi}^2/V \sigma^2_E$ indeed approaches unity.      

\begin{figure}
    \centering
    \begin{minipage}{.48\textwidth}
        \centering
        \includegraphics[width=1.2\linewidth]{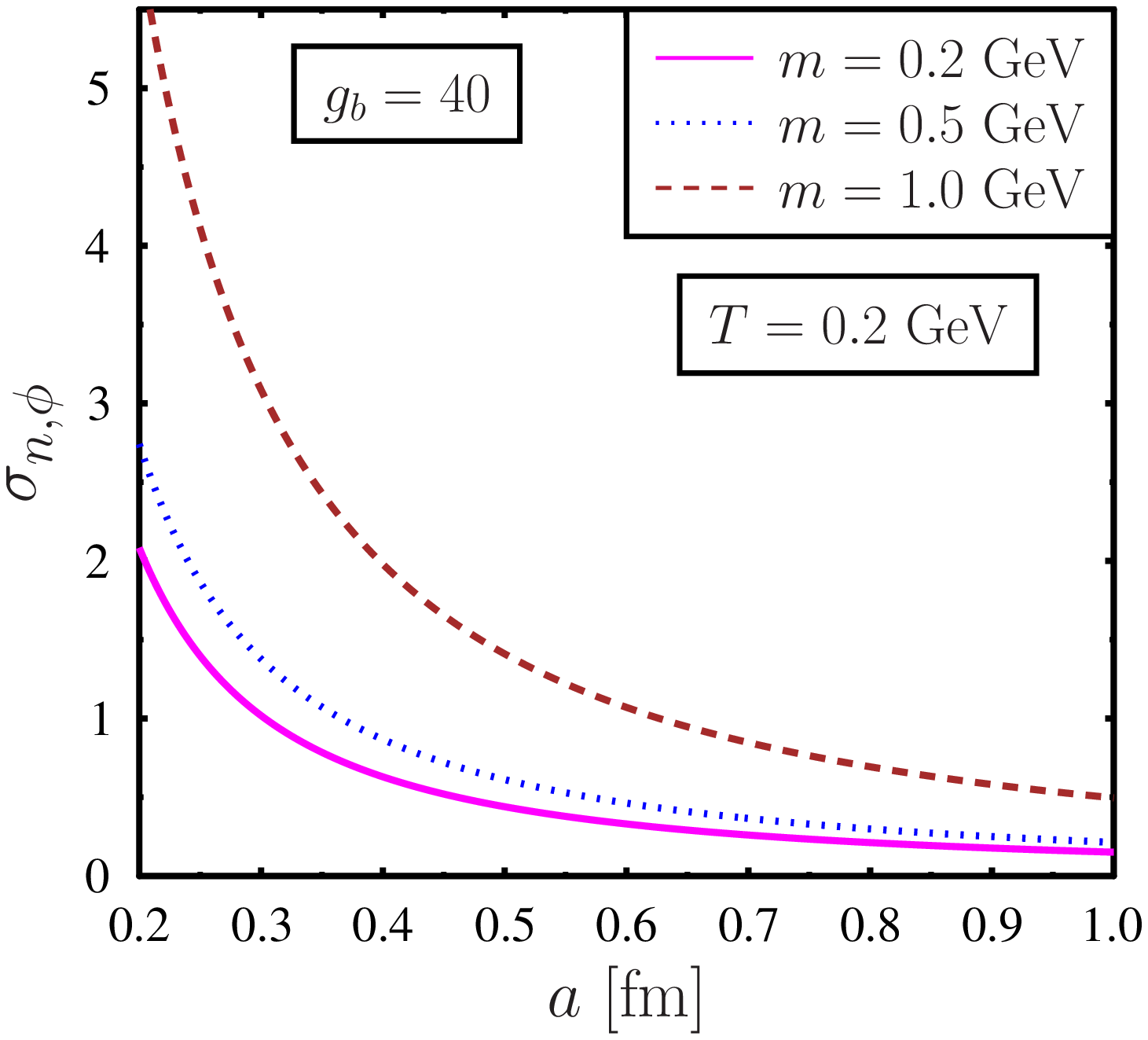}
    \end{minipage}~~
    \begin{minipage}{0.48\textwidth}
        \centering
        \includegraphics[width=1.2\linewidth]{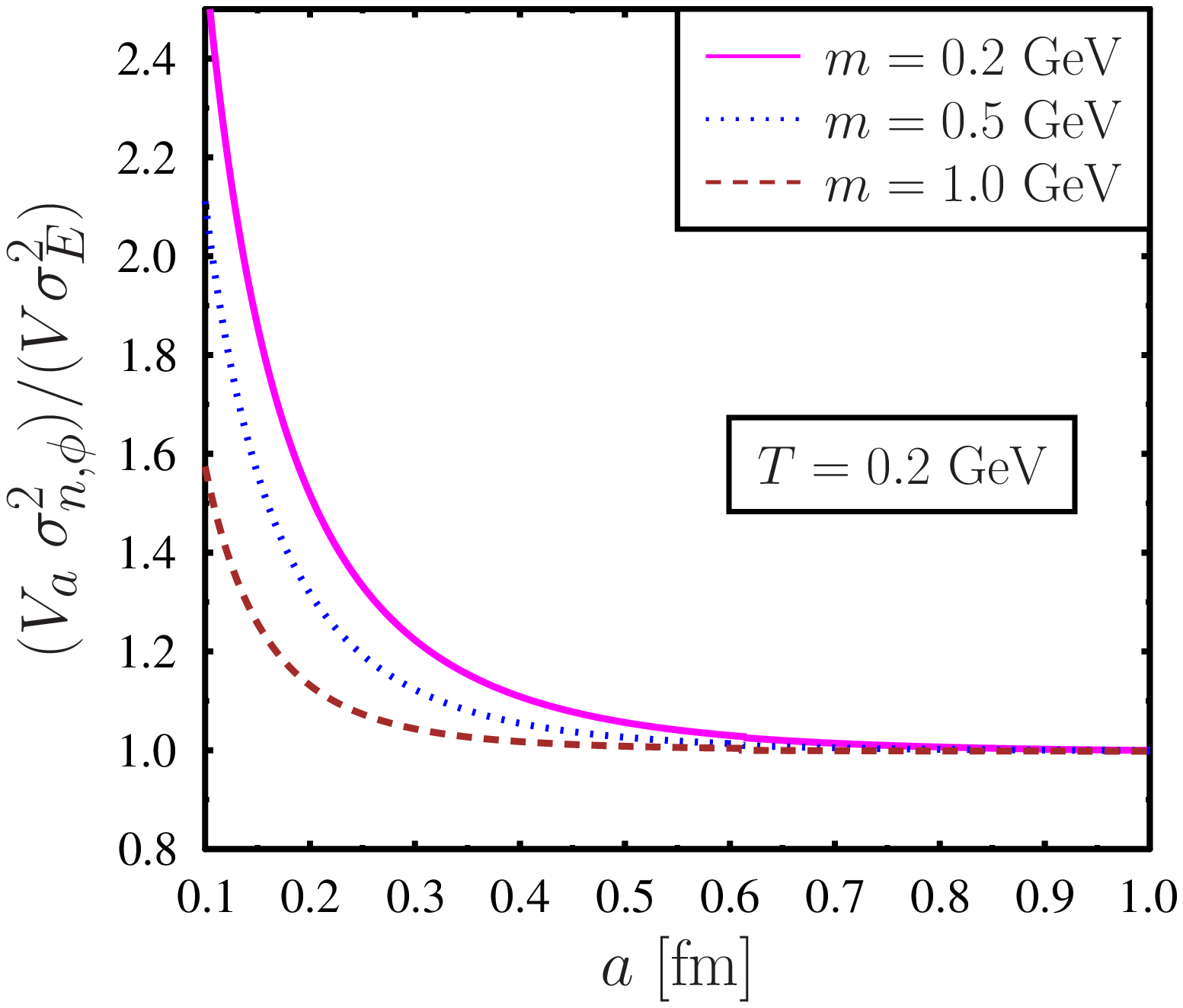}
    \end{minipage}
    \caption{Left plot: variation of normalized fluctuation $\sigma_{n,\phi}$ within the subsystem $\mathcal{S}_a$ described by a real scalar field with the scale `$a$' for different mass $(m)$ but fixed temperature $(T)$. Right plot: in this plot we show  the large system size limit by showing the variation of the normalized quantity $V_a \sigma_{n,\phi}^2/V \sigma^2_E$ with the scale `$a$' for different mass $(m)$ but fixed temperature $(T)$. For more discussions see text and Ref.~\cite{Das:2021acta}.}
\label{fig3}
\end{figure}

In the left plot of Fig.~(\ref{fig3}) we again show the variation of normalized fluctuation $\sigma_{n,\phi}$  with the scale `$a$' but for different mass $(m)$ for a fixed temperature $(T)$. From this plot, we can conclude that with an increase in mass ($m$) normalized fluctuation $\sigma_{n,\phi}$ increases. Here also we emphasize that the variation of $\sigma_{n,\phi}$ with mass is opposite to the variation of $\sigma_{\phi}$ ( not shown here). We also find that for $a\rightarrow 0$ fluctuation diverges and with the system size `$a$' quantum fluctuation decreases. In the right plot of Fig.~(\ref{fig3}) we show the variation of  the normalized quantity $V_a \sigma_{n,\phi}^2/V \sigma^2_E$ with the scale `$a$' for different mass ($m$) but fixed temperature $(T)$. From this plot, we observe that in the large volume limit or in the thermodynamic limit $V_a \sigma_{n,\phi}^2/V \sigma^2_E$ approaches unity which is consistent with the thermodynamic considerations. From Figs.~(\ref{fig2}) and ~(\ref{fig3}) we can conclude that the quantum statistical fluctuations become very important at a small length scale (say $a\sim 0.1$ fm). This indicates that at such a small length scale one can not naively use classical concepts of well-defined energy density. Rather one also has to take into consideration possible quantum effects.

\begin{figure}
    \centering
    \begin{minipage}{.48\textwidth}
        \centering
        \includegraphics[width=1.2\linewidth]{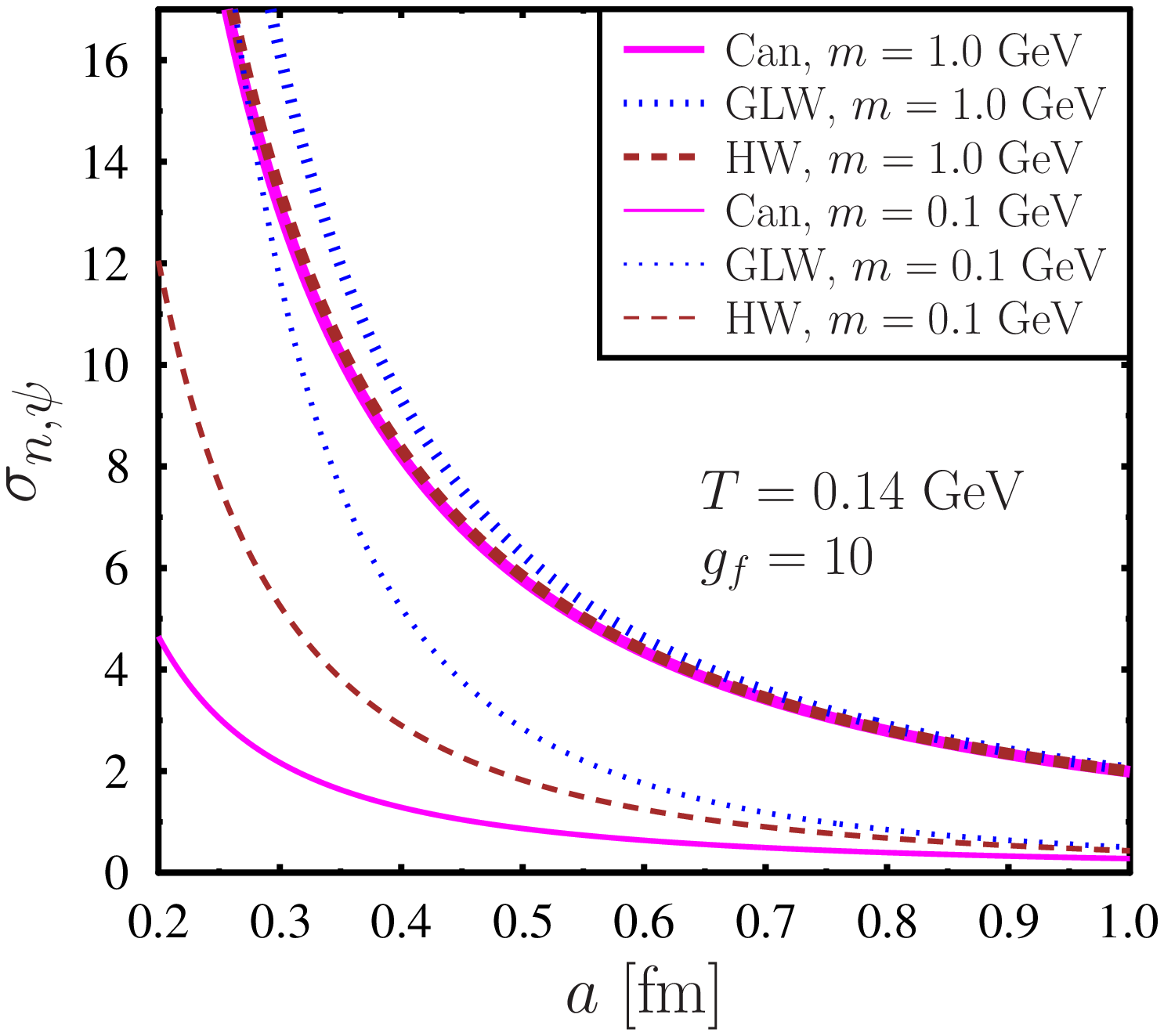}
    \end{minipage}~~
    \begin{minipage}{0.48\textwidth}
        \centering
        \includegraphics[width=1.2\linewidth]{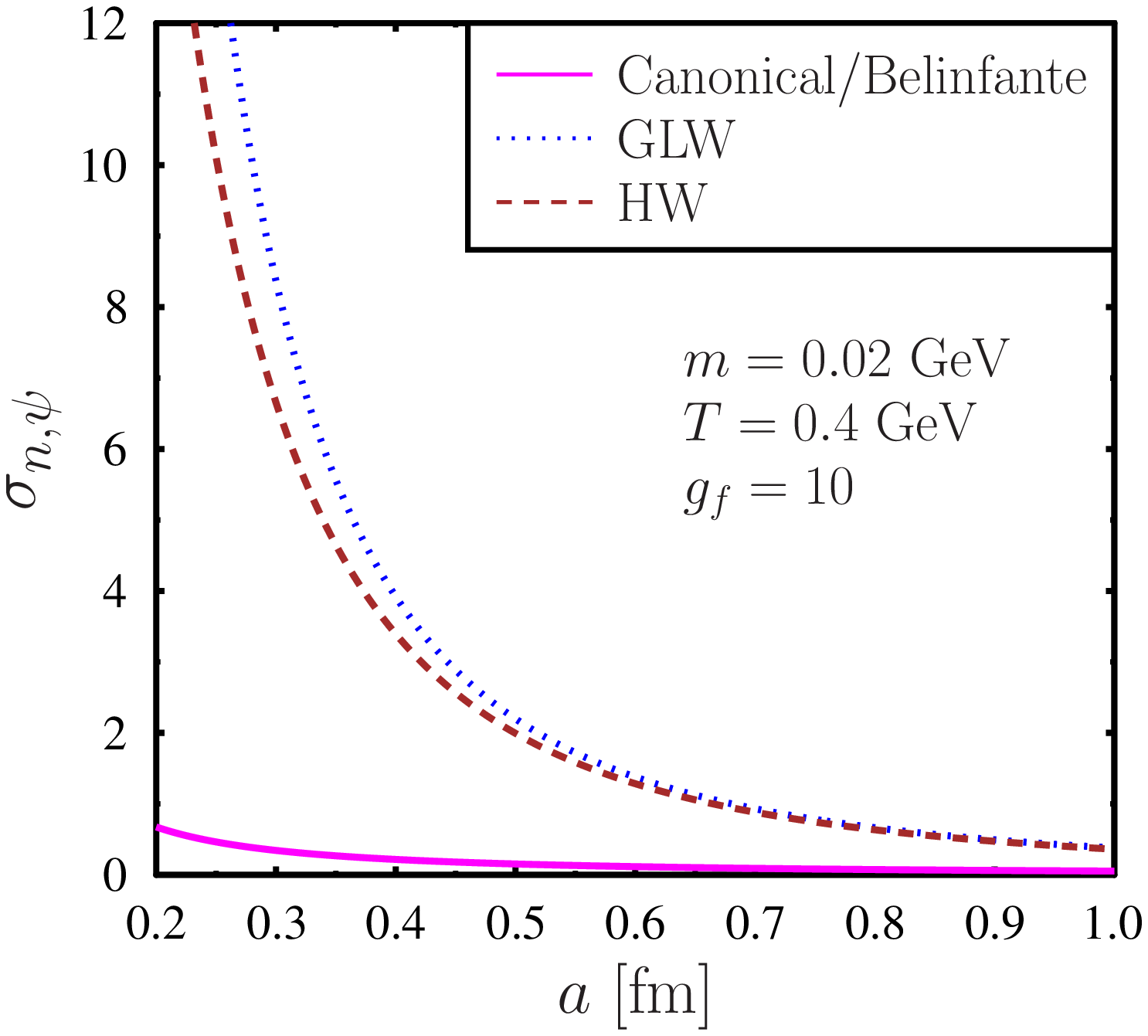}
    \end{minipage}
    \caption{Left plot: we show a comparison of $\sigma_{n,\psi}$ as obtained for different pseudo-gauge choices of the energy-momentum tensor for $T=0.14$~GeV but for two different masses. Right plot: we show a comparison of $\sigma_{n,\psi}$ for small mass and high temperature as obtained for different pseudo-gauge choices. Note that $\sigma_{n,\psi}$ is smallest for the canonical framework among various pseudo-gauge choices. For more discussions see text and Ref.~\cite{Das:2021aar}.}
\label{fig4}
\end{figure}

Once we have demonstrated the variation of quantum statistical fluctuation with the size of the sub-system, temperature, and mass, now we explicitly show the pseudo-gauge dependence of the quantum statistical fluctuation using various forms of the energy-momentum tensor for a fermionic field. In the left plot of Fig.~(\ref{fig4}) we show a comparison of the normalized standard deviation ($\sigma_{n,\psi}$) as obtained using different forms of the time-time component of energy-momentum tensor for $T=0.14$~GeV but for two different masses. In this plot the thick lines correspond to $m=1.0$~GeV and $T=0.14$~GeV. On the other hand thin lines correspond to $m=0.1$~GeV and $T=0.14$~GeV. From this plot, it is evident that for small values of $a < 0.6$~fm, the value of $\sigma_{n,\psi}$ differs with various pseudo-gauge choices. This difference in the value of $\sigma_{n,\psi}$ as obtained for various pseudo-gauge choices increases for smaller values of $a$. It is also interesting to note that for $m=1.0$ GeV the differences between different pseudo-gauges are smaller as compared to the case of $m=0.1$ GeV. In the right plot of Fig.~(\ref{fig4}) we show the variation of $\sigma_{n,\psi}$ for low mass ($m=0.02$ GeV) and high temperature ($T=0.4$~GeV). From this plot, one may conclude that with growing system size ($a$) the normalized standard deviation of fluctuations ($\sigma_{n,\psi}$) decreases for all pseudo-gauge choices. These figures also show that fluctuations crucially depend on the ratio $m/T$.

\begin{figure}
    \centering
    \begin{minipage}{.48\textwidth}
        \centering
        \includegraphics[width=1.2\linewidth]{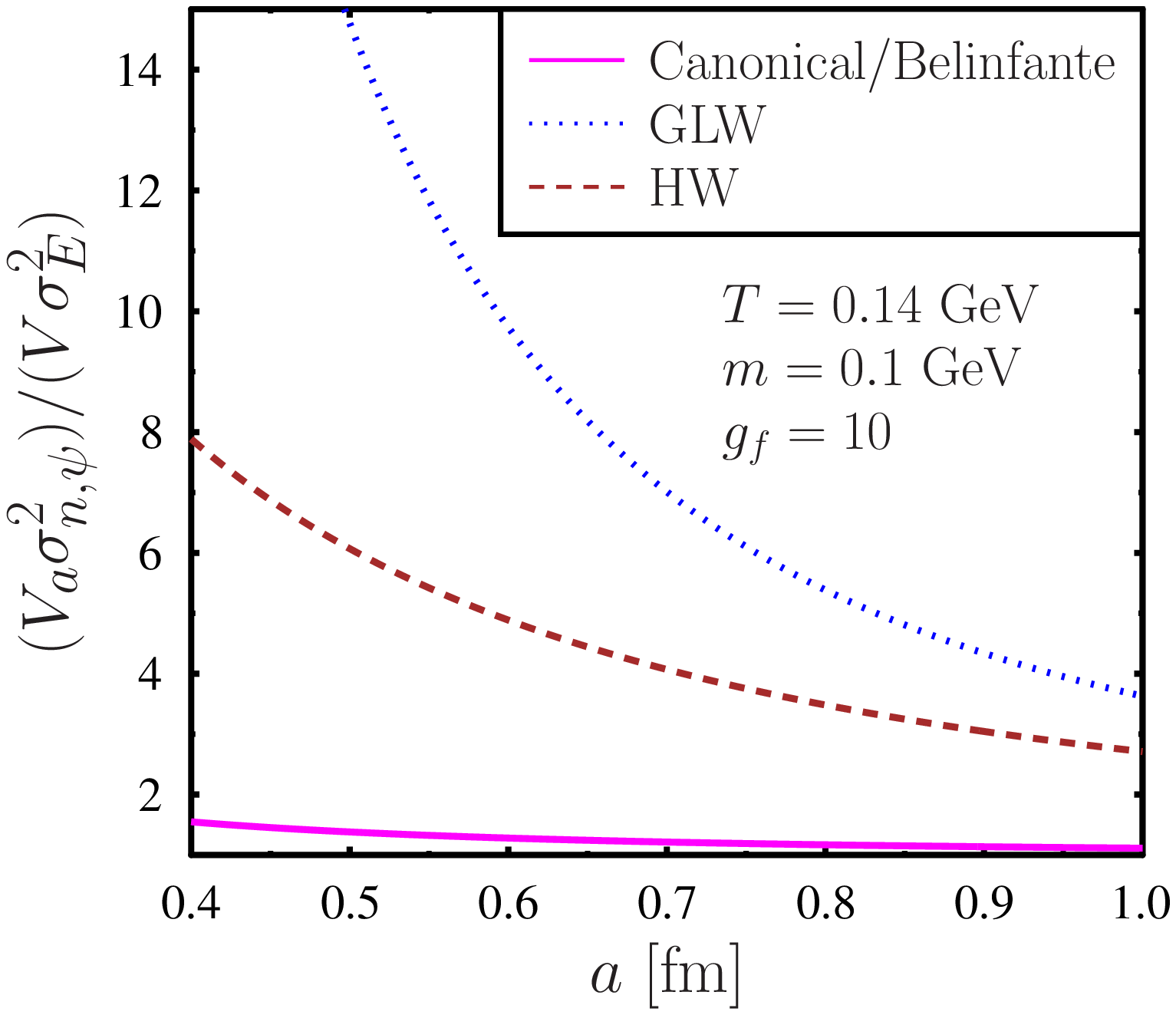}
    \end{minipage}~~
    \begin{minipage}{0.48\textwidth}
        \centering
        \includegraphics[width=1.2\linewidth]{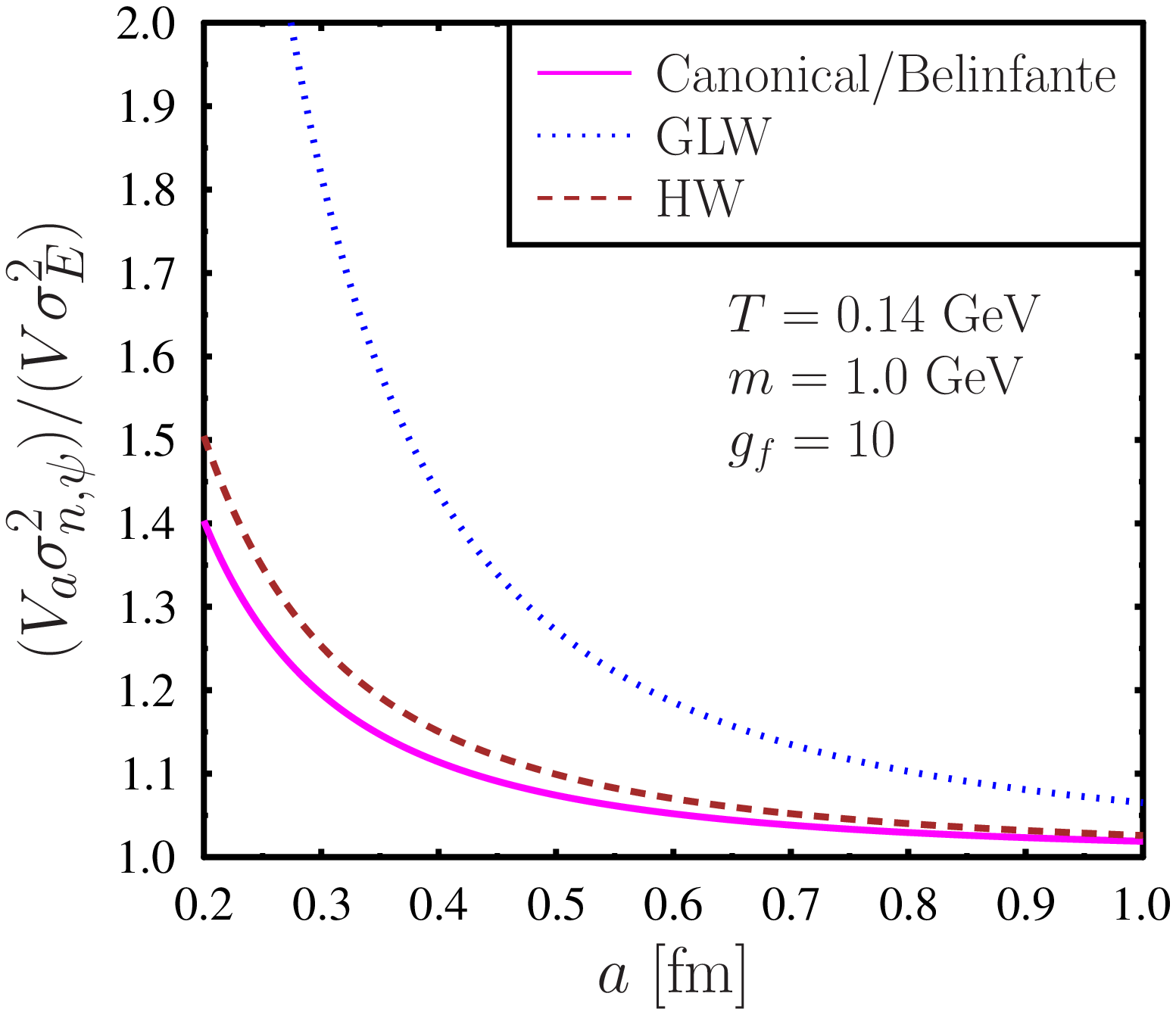}
    \end{minipage}
    \caption{Left plot: variation of the normalized fluctuation $V_a \sigma_{n,\psi}^2/V \sigma^2_E$ in the subsystem $\mathcal{S}_a$ with the length scale $a$ for $T=0.14$~GeV and $m=0.1$~GeV. Right plot: variation of $V_a \sigma_{n,\psi}^2/V \sigma^2_E$ in the subsystem $\mathcal{S}_a$ with the length scale $a$ but for $T=0.14$~GeV and $m=1.0$~GeV. From these plots we observe that in the large system size limit $V_a \sigma_{n,\psi}^2/V \sigma^2_E$ approaches unity. Note that $V_a \sigma_{n,\psi}^2/V \sigma^2_E$ is smallest for the canonical framework among various pseudo-gauge choices and it also approaches unity faster with the scale $a$. For more discussions see text and Ref.~\cite{Das:2021aar}.}
\label{fig5}
\end{figure}

Finally in Fig.~(\ref{fig5}) we demonstrate that in the large system size limit the dimensionless measure of the fluctuation $V_a \sigma_{n,\psi}^2/V \sigma^2_E$  approaches unity. In the left plot and in the right plot of Fig.~(\ref{fig5}) we show the variation of $V_a \sigma_{n,\psi}^2/V \sigma^2_E$ with the scale $a$ for $T=0.14$ GeV but for $m=0.1$ GeV and $m=1.0$ GeV respectively. From this plots one may observe that  among various choices of pseudo-gauge, for the canonical framework the decrease of $V_a \sigma_{n,\psi}^2/V \sigma^2_E$ with the scale $a$ is fastest. Also note that with increasing $m/T$
ratio, the decrease of $V_a \sigma_{n,\psi}^2/V \sigma^2_E$ with the scale `$a$' is faster.

\section{Conclusions}
\label{conclusion}
In this article, we have addressed the system size scaling of quantum statistical fluctuation by considering a gas of spin-half as well as spin-zero particles at finite temperature. We derived the formula for the quantum statistical fluctuations of energy in subsystems of a relativistic gas at finite temperature. A novel feature of such quantum statistical fluctuations is that they depend on the form of the energy-momentum tensor. However, for a sufficiently large system, different pseudo-gauges give rise to the same expression for the fluctuations which agree with the canonical-ensemble formula for energy fluctuation. Using explicit field theoretical calculations we argued that irrespective of different characteristics of the systems, i.e. bosonic or fermionic, and also irrespective of different pseudo-gauge choices a generic feature of quantum fluctuations is that they decrease with the system size and for a small length scale such fluctuations can be significant. On a practical side, our calculations can be used for quantitative estimation of fluctuations for given values of $T$, $m$, and degeneracy factor. If the dimensionless measure of the quantum fluctuations (say, $\sigma_n$) are small (smaller than 1) then only the classical picture of a well-defined energy density can be considered safely. These analyses also give us a practical way to determine the fluid cell or ``coarse-graining" size. 
If the ``coarse-graining" scale is small so that we can not neglect the quantum effects then quantum fluctuation should be combined in future works with hydrodynamic fluctuations~\cite{Kapusta:2011gt}. Moreover, these results can also be useful to determine a scale of coarse-graining for which the pseudo-gauge choice becomes irrelevant in the context of hydrodynamic modeling of high-energy collisions. Using these results, for a given value of $T$, $m$ and the degeneracy factor one can determine the scale $a_0$ so that above the scale $a_0$ fluctuations are small and pseudo-gauge independent. Then such a scale $a > a_0$ can be safely used as the ``coarse-graining" scale or scale for the fluid element. What remains rather non-trivial to interpret is the pseudo-gauge dependence of fluctuations for small systems. From an experimental point of view, one can measure the expectation values of some combination of the quantum field theoretical operators. However we have argued that such expectation value can depend on the form of the operator or pseudo-gauge. Any realistic measurements of such pseudo-gauge dependent quantities need to be explored rigorously~\cite{Becattini:2012pp,Nakayama:2012vs}. Moreover these analysis
contains important results which may be relevant to a broader scientific community,
in particular with regards to the increased interest in small systems produced
in heavy ion collisions~\cite{Kurkela:2019kip}. 

%%%%%%%%%%%%%%%%%%%%%%%%%%%%
\section*{Acknowledgments}
I would like to thank the organizers of the ``10th International Conference on New Frontiers in Physics (ICNFP 2021) " for giving me an opportunity to present our recent works. I thank Wojciech  Florkowski,  Radoslaw  Ryblewski,  and  Rajeev  Singh for a fruitful collaboration. My research work is supported by the Polish National Science Centre Grant No. 2018/30/E/ST2/00432. 
%%%%%%%%%%%%%%%%%%%%%%%%%%%%

\section*{References}

\bibliographystyle{utphys}
%\bibliography{jetquenching}
\bibliography{icnfp_proc2019}

\end{document}